
\def\d#1/d#2{ {\partial #1\over\partial #2} }



\def\Tr{\hbox{Tr}}


\newcount\eqnumber
\def\beq{ \global\advance\eqnumber by 1 $$ }
\def\eeq{ \eqno(\the\eqnumber)$$ }
\def\n{\global\advance \eqnumber by 1\eqno(\the\eqnumber)}
\def\puteqno{
\global\advance \eqnumber by 1 (\the\eqnumber)}
\def\beqs{$$\eqalign}
\def\eeqs{$$}


\def\ifundefined#1{\expandafter\ifx\csname
#1\endcsname\relax}
 \newcount\sectnumber \sectnumber=0
\def\sect#1{ \advance \sectnumber by 1 {\it \the \sectnumber. #1} }

\newcount\refno \refno=0  
\def\[#1]{
\ifundefined{#1}
\advance\refno by 1
\expandafter\edef\csname #1\endcsname{\the\refno}\fi[\csname
#1\endcsname]}
\def\refis#1{\noindent\csname #1\endcsname. }

\def\label#1{
\ifundefined{#1}
\expandafter\edef\csname #1\endcsname{\the\eqnumber}
\else\message{label #1 already in use}
\fi{}}
\def\(#1){(\csname #1\endcsname)}
\def\eqn#1{(\csname #1\endcsname)}

\parskip=12pt
\magnification=1200
\def\BEGINIGNORE#1ENDIGNORE{}

\baselineskip=10pt




                                             \hfill UR-1388
                                             \hfill ER40685-837

\centerline{ \bf Poisson Algebra of Wilson Loops in Four--dimensional
Yang--Mills Theory}

\vskip.1in
\centerline{\bf    }
\vskip.1in
\centerline{\rm   S. G. Rajeev and  O.T. Turgut}
\vskip.1in
\centerline{\it Department of Physics and Astronomy}
\centerline{\it University of Rochester}
\centerline{\it Rochester, N.Y. 14627}
\centerline{\it e-mail: rajeev and
 turgut@urhep.pas.rochester.edu}
\vskip.4in

\baselineskip=24 true pt

\centerline{\bf Abstract}

We formulate the canonical structure of Yang--Mills theory in terms of Poisson
brackets of gauge invariant observables analogous to Wilson loops.
This algebra is non--trivial and tractable in a light--cone formulation.
For $U(N)$
gauge theories the result  is a Lie algebra while for $SU(N)$ gauge theories it
is a quadratic algebra.  We also study the identities satsfied by the gauge
invariant observables. We suggest that the phase space of  a Yang--Mills
theory is a co--adjoint orbit of our Poisson algebra; some partial results in
this direction are obtained.

\vfill\break

\sect{ Introduction}

A central problem in present day particle physics is to find a formulation of
Yang--Mills  theories in terms of manifestly gauge invariant variables. This
problem was tackled first by Mandelstam \[mandelstam]. It is even more
important to solve this problem now since there is strong experimental evidence
that in an unbroken gauge theory ( Quantum ChromoDynamics, QCD) all observable
states (
hadrons) are gauge invariant. Thus a theory of hadrons would be  such a
reformulation of QCD in terms of gauge invariant variables.
 This problem has been solved in two dimensions,
for pure gauge theory,\[rajeev]  for gauge theory  coupled to
non--relativistic matter \[turgutetal],\[chandar],
and coupled to Dirac fermions
\[trieste],\[twodbaryon],\[twodhadron] ( For related work, see
\[pesando],\[wadia].) In all these cases it was possible to cast the
canonical formalism of the theory in manisfestly gauge invariant form by an
appropriate choice of variables.

Often the resulting phase space has little resemblance to the original one. In
the case of two dimensional QCD, the phase space of the gauge invariant
formalism is a Grassmannian. The dynamical variable is a function of two points
of space--time ( which lie on a null line), $M(x,y)$ .This variable must
satisfy a
quadratic constraint. ( The set of such `matrices' satisfying this quadratic
constraint  is a coset space of the Unitary group, the Grassmannian.)
The Poisson brackets of this variable follow from the natural symplectic
structure of the Grassmannian. The hamiltonian is a quadratic function of
$M(x,y)$. These results were originally derived by rewriting the large $N_c$
limit of two dimensional QCD as a classical dynamical system \[trieste].
In two dimensional QCD, $M(x,y)$ has the meaning of a quark bilinear.
The theory of the variable $M(x,y)$ is thus a theory of hadrons: mesons are the
small oscilations around the vacuum and baryons are the solitons \[twodbaryon].
It was even possible to reverse the whole argument and derive two dimensional
QCD as the quantization of the  hadron theory.
A similar formulation of QCD of systems with spherical symmetry has also been
possible \[sphqcd].

A  similar hadron theory  in  four dimensions  is  not yet possible.
In this paper we take some steps in that direction: in some sense the
`kinematical' part of the problem is solved. The idea was  outlined earlier in
ref. \[trieste]. Following the strategy that  worked in two dimensional QCD, we
first find the canonical commutation relations of a set of gauge invariant
variables. The natural choice for gauge invariant variables are the Wilson
loops, the trace of the parallel transport operators around closed loops.(
There is a large amount of work on the loop formulation of gauge theories. A
partial list is
Ref.\[mandelstam],\[migdal],
\[giles],\[kikkawa],\[ashtekaretal],\[annphys].)
In the usual canonical formalism, these variables will have a trivial Poisson
algebra if the loops lies on a space--like initial value surface: the spatial
components of the connection are the `position' variables of the Yang--Mills
phase space. To encode the  canonical information, we must introduce some gauge
invariant `strip' variables
from the electric field \[ashtekaretal].

 We find it more natural, and more analogous to the approach in two
dimensions to prescribe initial data on a null surface. Then  we can eliminate
the non--propagating, time like, component of the connection   explicitly.
( Light--frame formulation of QCD has attracted much attention recently for
independent reasons \[lightcone].)
Moreover, the  spatial components are
canonically conjugate to each other. The Wilson loops on the null surface have
non--trivial Poisson brackets and completely encode the canonical information
of the Yang--Mills phase space. In fact the algebra so obtained is quite
elegant; it has a simple description in terms of the geometry of the loops.
 It is  a Lie algebra in the case where the underlying  gauge group is
$U(N_c)$.
\[trieste]. In the realistic case of an $SU(N_c)$ gauge theory, it is a
quadratic algebra. We believe it is the first time that such a quadratic
algebra has
emerged in a physically relevant theory. (The Wilson loops of QCD  will satisfy
this quadratic algebra.) The algebra  has a particularly simple form for smooth
loops in a {\it four--dimensional} gauge theories; it is not well--defined for
dimensions higher than four and is somewhat more complicated in dimension
three.

In the case of two dimensional QCD, the phase space of the hadron theory was a
co--adjoint orbit of the Poisson algebra of observables. We conjecture that the
Phase--space of a  four dimensional Yang--Mills theory is a co--adjoint orbit
of our Poisson algebra. The quantum analogue of this statement is that
different    Yang--Mills theories correspond  to  different  representations (
or deformations ) of the
same underlying  Poisson algebra. We are able to show this for the simplest
case of
$U(1)$ gauge group.  In this case the Yang--Mills phase space is given by a
quadratic constraint ( Mandelstam constraint) on the Wilson loop  variable.
In more complicated cases the Yang--Mills phase space is given by more
complicated Mandelstam constraints; we will see that these are invariant
subspaces of the Poisson algebra. We  have not shown that the action of the
Poisson algebra is transitive, but we believe this to be the case.

 The conformal invariance of  four dimensional Yang--Mills theory allows us to
map the problem to the conformal completion of Minkowski space; then every loop
must pass through the point representing past infinity,
which makes it more convenient to state the Mandelstam constraints.

It is of much interest to understand the representation theory and the
automorphisms of our algebras. Also, the algebra might possess interesting
quantum deformations in the sense of Drinfeld \[drinfeld]. This should help in
developing the classical ideas of this paper into the quantum regime.
Just as two dimensional hadron theory is a  field theory of open strings, the
algebra we describe here should be the basis of a four dimensional  closed
string field theories. It is not nilpotent  unlike the light--cone string field
algebra of conventional string theories which are just canonical commutation
 relations \[kakukikkawa]. Some of the
nonlinearities of the theory are contained in the Poisson algebra, just as in
two dimensional hadrondynamics. We hope that the string field theoretical
aspects of the subject will be investigated further in the future.

We first give a canonical formulation of Yang--Mills theory in a particular
kind of
light--cone co--ordinates.  We also describe how using conformal invariance the
problem can be mapped from Minkowski space to its conformal completion.
Then we derive the Poisson algebra of the Wilson loops. We note that in
four--dimensional Yang-Mills theory the algebra can be simplified. Then we
discuss the Mandelstam identities satisfied by the gauge invariant variables.
We believe that the Poisson algebras we find are of independent interest ( for
example as  a starting point for closed string field theory); therefore we give
a direct proof of the Jacobi identity of the algebra without using its
representation in terms of gauge fields. We show that the Mandelstam identities
describe surfaces in the dual of the Lie algebra that are invariant under the
action of the algebra. Many of the calculations and technical details are in
the appendices.

\sect{Yang-Mills theory in the light-cone coordinates and the symplectic
form }

In this section we will give a formulation of Yang-Mills theory using
light cone coordinates\[lightcone]. The advantage of this formalism is that the
Yang-Mills action is first order in its "time" derivatives and thus
the Yang-Mills potential becomes a  self-conjugate variable  in the Hamiltonian
formalism. This makes it possible to describe the canonical structure entirely
in terms of the gauge invariant loop variables constructed out of the
Yang--Mills potential.

We will first state the conventions that will be
used in $d$ space-time dimensions.We define the light-cone variables as
follows;
\beq
      u={t-r} \qquad   r,
      \theta^1, \theta^2, \dots ,\theta^{d-2}
.\eeq
where $t$ and $r$ are the time and radial coordinates in Minkowski
space and $\theta$'s are the corresponding angular variables in $d$
dimensions. If one uses these coordinates, the metric can be expressed
as,
\beq
    ds^2=du(du+2dr)-{r^2}((d\theta^1)^2 + \sin^2\theta^1 (d\theta^2)^2 +
  \dots + \sin^2\theta^1\sin^2\theta^2\dots
      \sin^2\theta^{d-3}(d\theta^{d-2})^2)
.\eeq
In this paper $d=4$ will be of special interest and we will state
the metric in more conventional coordinates;
\beq
    ds^2=du(du+2dr)-{r^2}(d\theta^2 + \sin^2\theta d\phi^2)
.\eeq
It is useful to find some shorthand symbols for the angular
scaling factors in the metric and we define accordingly
$\Phi_1, \Phi_2 \dots \Phi_{d-2}$ as,
\beq
   \Phi_1=1, \quad \Phi_2=\sin\theta^1,\quad
    \dots \Phi_{d-2}=\sin\theta^1\sin\theta^2...
    \sin\theta^{d-3}.
\eeq
Also, $\surd{-g}=r^{d-2}\Phi$, where
$\Phi= \sin^{d-3}\theta^1 \sin^{d-4}\theta^2\dots \sin\theta^{d-3}$.
Yang-Mills theory is defined through the vector potential $A_\mu$,
which is a Lie algebra valued vector. The Yang-Mills action is
given by,
\beq
    S=-{1\over 4} \int \surd{-g} {\rm Tr} F_{\mu\nu}F^{\mu\nu}
    dudrd\Omega.
\eeq
where $d\Omega=d\theta^1\dots d\theta^{d-2}$ and
$F_{\mu\nu}=\partial_\mu A_\nu-\partial_\nu A_\mu + [A_\mu,A_\nu]$.
$A_\mu$ can be written as $A_\mu^a T_a$ where $T_a$ are the generators
of the Lie algebra $\underline G$. We will usually take the structure
group $G$ to be $U(N)$, although the case of $SU(N)$  will be occassionally
discussed.

 With our  choice of variables,  $\partial \over \partial u$
 is in fact a time like vector and $\partial \over \partial r$ is null.
We will regard $u$ as the evolution variable so that the initial
data is given on the light-cone $u=\rm constant$. It will be convenient to
choose the gauge $A_r=0$.
As a  result,
\beqs{
      F^{ur}=F_{ru},\quad F^{ri}={1 \over r^2}\Phi_i^{-2}(F_{ri}-F_{ui}),\quad
       F^{ui}=-{1 \over r^2} \Phi_i^{-2}F_{ri},\quad
       F^{ij}={1 \over r^4} \Phi_i^{-2} \Phi_j^{-2} F_{ij}
.}\eeqs
Also,
\beq
    F_{ui}=\partial_u A_i-\nabla_i A_u,\quad F_{ri}=\partial_r A_i,\quad
    F_{ur}=-\partial_r A_u,\quad
    F_{ij}=\partial_i A_j-\nabla_j A_i
.\eeq
where we defined the covariant derivative as $\nabla_i=\partial_i+ [A_i,\  ]$.

\beqs{
         S=-{1\over{4\alpha^2}} \int &\Phi r^{d-2}
     \{-2\Tr (\partial_rA_u)^2 -4r^{-2}  \Phi_i^{-2}
       \Tr (\partial_u A_i\partial_r A_i) + \cr
           &4r^{-2} \Phi_i^{-2}
       \Tr(\partial_r A_i \nabla_i A_u) +
       2r^{-2}\Phi_i^{-2} \Tr (\partial_r A_i \partial_r A_i)\cr
       &+ r^{-4} \Phi_i^{-2} \Phi_j^{-2} \Tr F_{ij}F_{ij}\}dudrd\Omega
.}\eeqs
where we assumed summation over repeated indices.

Since $A_u$ does not have a "time" derivative ( i.e., derivative with respect
to the evolution variable $u$ ) in the action we may eliminate
it by using its equation of motion and get a reduced action:
\beq
      \Phi \partial_r(r^{d-2}\partial_r A_u)-r^{d-4}\nabla_i(\Phi_i^{-2}
      \partial_r A_i)=0
.\eeq
Using the integral operator $\partial_r^
{-1}$ we get,
\beq
     A_u= \Phi^{-1} \partial_r^{-1}{1\over {r^{d-2}}}\partial_r^{-1}
      ( r^{d-4}\nabla_i(\Phi_i^{-2} \partial_r A_i))
.\eeq
This equation is enough to eliminate $A_u$ and obtain the reduced
action as,
\beqs{
       S={-1 \over 4\alpha^2}& \int \{ -2r^{2-d} \Phi^{-1} \Tr
          (\partial_r^{-1} r^{d-4} \nabla_i (\Phi_i^{-2} \partial_r
           A_i))^2- 4 r^{d-4} \Phi \Phi_i^{-2} \Tr \partial_u
           A_i \partial_r A_i \cr
             &+4r^{d-4} \Phi \Phi_i^{-2} \Tr (\nabla_i \Phi^{-1}
              \partial_r^{-1} r^{2-d} \partial_r^{-1} r^{d-4}
            \nabla_j(\Phi_j^{-2}\partial_r A_j))(\partial_r A_i) \cr
             &+2 r^{d-4} \Phi  \Phi_i^{-2} \Tr \partial_r A_i \partial_r A_i
              + r^{d-6}\Phi \Phi_i^{-2} \Phi_j^{-2} \Tr F_{ij}F_{ij}\}
                dudrd\Omega
.}\eeqs
This action is already first order in its evolution variable. Therefore
$A_i$ on the initial surface $u=$ constant, are co--ordinates on  the phase
space of the theory.
We can read off the Poisson brackets of these  variables from the above action.
The anti--symmetric part of the operator $\Phi \Phi_i^{-2}
r^{d-4}\partial_r$ (putting $\alpha=1$ for simplicity) can be interpreted
to be the  the symplectic form. The Poisson brackets are given by its inverse.
Let us define the Green's function $f(r,r')=<r|2(r^{d-4}\partial_r + \partial_r
r^{d-4})^{-1}|r'>$
and impose  the antisymmetry condition, $f(r,r')=-f(r',r)$. We can solve  its
defining differential equation to get,
\beq
     f(r,r')= {1 \over 2}(rr')^{2-d/2}{\rm sign}(r-r')
.\eeq
Here we define sign function to be,
\beq{
       {\rm sign}(r-r')=\cases{ 1, &if $r>r'$ \cr
                                0, &if $r=r'$ \cr
                                -1, &if $r<r'$.\cr}
}\eeq

Now we can write down the Poisson brackets of the variables $A_i$:
\beq
      \{ A_i^a (\theta^k,r,u), A_j^b (\theta'^k, r',u') \}=\delta^{ab}
        \delta_{ij} \delta^{d-2}(\theta^k-\theta'^k)\Phi^{-1}\Phi_i^2
        f(r,r')
.\eeq

It is most natural to choose the initial value surface $u=$ constant to be the
light-cone in the infinite past, since,  all the dynamical degrees of
freedom are radiative.
This will also enable us to express the canonical structure above  more
elegantly in terms of gauge invariant variables. However, in the co--ordinate
system we have been using, this will be a  singular limit. One solution is to
make a conformal mapping of the metric tensor so that the light--cone in the
infinite past is brought to a finite `distance'.  For $d=4$, classical
Yang--Mills theory  is conformally invariant, so that we can  use this
conformal mapping quite naturally. We will see  several other
simplifications in the four dimensional case. It is gratifying that these
simplifications occur in the physically interesting case.

\sect{ Yang--Mills Theory on the Conformal Completion of the Minkowski Space}

Let us recall briefly the conformal completion of Minkowski space.
The idea is to bring the points at infinity to a finite distance by using
a conformally equivalent metric:
\beq{
          g_{\mu \nu} \mapsto \hat g_{\mu \nu}=\Omega^2 g_{\mu \nu}
.}\eeq
Moreover new co--ordinates are found, which remain finite
at these `points at infinity'. We will refer
to \[penrose] for the details. If we denote the Minkowski space by $\cal
M$ then it is possible to adjoin
to $\cal M$ a certain boundary surface, called $\cal I$, in such a way that
$\hat g_{\mu \nu}$ extends smoothly to this boundary. This new space
is denoted by $\bar {\cal M}$ and we have $\cal I={\partial}$$ \bar {\cal M}$
and
${\cal M}={\rm int}\bar {\cal M}$. We define
\beq
       u=t-r \quad {\rm and }\quad  v=t+r
.\eeq
We further introduce $R$ and $T$ as
\beq
          R=\tan^{-1} v-\tan^{-1} u \quad T=\tan^{-1} u +\tan^{-1} v
.\eeq
and define $U=T-R$. Then we can express the new metric in terms of
$U$ and $R$ and the angular  coordinates;
\beq
   d \hat s^2= dU(dU+2dR)-\sin^2R((d\theta^1)^2+\sin^2 \theta^1 (d\theta^2)^2)
.\eeq
This is conformally equivalent to the old metric with conformal factor
\beq
	\Omega=2(1+u^2)^{-1/2}(1+v^2)^{-1/2}.
\eeq

This space $\bar {\cal M}$  is usually represented by two cones attached at
their openings,
and the resulting spherical intersection  is identified to one point,
called $i^0$. This point represents the spatial infinity [Refer to
figure-1]. The conic boundaries are  denoted by  ${\cal I}^{-}$ and
${\cal I}^{+}$ respectively.

 If we define $V=T+R$,
the region  $- \pi < U < \pi$ and $- \pi < V < \pi$ corresponds to ${\cal M}$.
By extending their domains smoothly $U=-\pi$ corresponds to
${\cal I}^-$ and $V= \pi$ to ${\cal I}^+$.
 We also have past timelike infinity
$i^-$ given by  $U=V=-\pi$  and future timelike infinity $i^+$ given
by $U=V= \pi$. The spatial infinity $i^0$ corresponds to just one point
$U=-V=-\pi$. Physically ${\cal I}^-$ denotes the future light cone of
past timelike infinity and ${\cal I}^+$ denotes the past light cone of
future timelike infinity. (In the references they are called past null
cone and future null cone respectively, and this is in agreement with
with our terminology if we take $i^0$ to be the reference point).

 It is possible  also to  identify all three points
$i^-$, $i^+$ and $i^0$ to one point yielding a compact space--time, with
topology $S^3\times S^1$.  The ( trace of ) parallel transport operators
around around the homotopically non--trivial curves on this space will provide
gauge invariant obeservables for Yang--Mills theory. These are in fact the
variables we will use. We find it more convenient to not perform this
identification, so that the
 observables  we consider will be the ( trace of) parallel transport operators
along open curves from $i^-$ to $i^0$.

In 4 dimensions Yang-Mills field equations are conformally invariant. So the
phase space of Yang-Mills theory in Minkowski background can be mapped
to the space of solutions on an open domain of $S^3 \times R$ with
appropriate initial data. We will write the action in 4 dimensions using this
conformal mapping and give initial conditions on the future light cone of
past timelike infinity $\cal I^-$. In other dimensions one can
still specify initial data on the same surface and write down the Yang-Mills
theory on an analogous space, but there is no direct correspondence between
this theory and the Yang-Mills with Minkowski background. Therefore  such a
conformal map is most useful in 4 dimensions.

Since the only difference in the metric is the introduction of $\sin R$ instead
of $r$, we can read of the results immediately for the action by replacing
$r$ with $\sin R$. We are interested in the symplectic form, so we will
state the result directly;
\beqs{
       S=-{1\over 4} \int &\{-2 \sin^2 R \Phi^{-1} \Tr (\partial_R^{-1}
                      \nabla_i (\Phi_i^{-2} \partial_R A_i))^2
                     +4 \Phi \Phi_i^{-2} \Tr \partial_U A_i \partial_R A_i\cr
                  &+4 \Phi \Phi_i^{-2} \Tr \nabla_i \Phi^{-1} \partial_R^{-1}
                  \sin^{-2} R \partial_R^{-1} \nabla_j(\Phi_j^{-2}
                  \partial_R A_j) (\partial_R A_i) \cr
                     &+2 \Phi \Phi_i^{-2} \Tr \partial_R A_i \partial_R A_i
                  + \sin^{-2}R \Phi \Phi_i^{-2} \Phi_j^{-2}\Tr F_{ij} F_{ij}\}
                   dRdUd\theta^1 d\theta^2
.}\eeqs
and this gives,
\beq
     \{ A_i^a(R, \theta^k, U), A_j^b(R', \theta'^k, U) \}=
      \delta_{ij} \delta^{ab} \delta^2(\theta^k-\theta'^k)\Phi^{-1}
        \Phi_i^2 h(R,R')
.\eeq
where $h(R,R')={1 \over 2} {\rm sign} (R-R')$.

{}From now on for 4 dimensional Yang-Mills theory we will use this
Poisson bracket and the conformal completion of the Minkowski space.
For other dimensions we will continue to use the Minkowski background.

\sect{  Wilson ``Loops" and the Poisson Algebra}

We will start from the four dimensional case and give the corresponding
formulae in  other dimensions.
Once we impose  the gauge  condition $A_R=0$ the residual gauge transformations
have to be independent of $R$. Thus a parallel transport around a curve on a
$U=$ constant surface  will be gauge--invariant  as long as it is closed in the
 angular directions: it does not need to be closed in the radial direction.
We will still call them ``loops", as we cannot
 think of a better name. A complete
set of obserbles is provided by ``loops" that lie entirely on a $U=$ constant
surface. This is because  the $A_U$ component  carries no new information: it
can be eliminated in terms of $A_i$.

 As a result we define a   ``loop"
to be piecewise differentiable maps $\xi:[0,L] \to {\bar {\cal M}}$ such
that $\xi^U$ remains constant,  $\xi^i=\theta^i$ for $i=1,2$ and
$\xi^i(0)=\xi^i(L)$.

Define also the Wilson ``loop'' variable ,
\beq
W(A,\xi) \equiv \Tr {\hat e}^{\int_{\xi}  A_\alpha d\xi^\alpha}.
\eeq
We are considering $A_\mu$ as matrices in  the fundamental representation of
$U(N)$. Also,
$\hat e$ denotes the  path ordered exponential. From our earlier comments it is
clear that $W(\xi)$ form a complete set of gauge invariant observables.

We now wish to express the  symplectic structure of the Yang--Mills theory in
terms of these gauge invariant observables. The computation is in principle
straightforward. It turns out to be convenient to replace the curve $\xi$ by a
piece-wise linear approximation to it.
The details are given in Appendix-1. We get the following Lie algebra, first
obtained in Ref. \[trieste]:
\beqs{
     &\{ W(\xi_1), W(\xi_2) \} =\cr
             & \int d\theta^1 d\theta^2 ds_1 ds_2
        \delta^2(\xi_1^k(s_1)-\theta^k) \delta^2 (\xi_2^k(s_2)-\theta^k)
        h(R(s_1),R(s_2)) \xi_1'.\xi_2' W(\xi_1 {\circ}_\theta \xi_2)
.}\eeqs
where we defined $\xi_1'.\xi_2'$ to be $\sum_k {d\xi_1^k \over ds_1}
{d\xi_2^k \over ds_2} \Phi_k^2 \Phi^{-1}$ and
 also $W(\xi_1 {\circ}_\theta \xi_2)$
denotes a ``loop" product  we will explain below.

These  Poisson brackets of the Wilson `loops' form a Lie algebra only for  a
specific
normalization of the $U(1)$ charge:  when it has the same length as the $SU(N)$
generators  under the Killing
form. This is the value of the $U(1)$ coupling constant for which the large $N$
limit is well--defined\[thooft].For  the $SU(N)$ gauge theory
instead we would obtain the following result,
\beqs{
       \{ W(\xi_1), W(\xi_2)&\} = \int d\theta^1 d\theta^2
        ds_1 ds_2 \delta^2(\xi_1^k-\theta^k) \delta^2(\xi_2^k-\theta^k)
         h(R_1,R_2) \xi_1'. \xi_2' W(\xi_1 {\circ}_\theta \xi_2)  \cr
           &-{1 \over N}\int d\theta^1 d\theta^2 ds_1 ds_2
              \delta^2 (\xi_1^k-\theta^k) \delta^2 (\xi_2^k-\theta^k)
              h(R_1, R_2)  \xi_1'. \xi_2' W(\xi_1) W(\xi_2)
.}\eeqs
 which is a quadratic Poisson
algebra   and  not  a Lie Algebra(we suppressed the explicit s dependence).
 Thus from this point of view the $U(N)$
gauge theory with $U(1)$ charge  chosen as above is simpler than an $SU(N)$
gauge theory.

This algebra is invariant under reparametrization of the curves, $s\to \hat
s=\phi(s)$, $\phi$ being monotonic. If the curve future pointing everywhere,
$R(s)$ is a monotonic function, and we can use $R$ itself as a parameter.
That would reduce the above expression to
\beq
    \{ W(\xi_1), W(\xi_2) \} =  \int d\theta^1 d\theta^2
        dR_1 dR_2 \delta^2(\xi_1^k-\theta^k) \delta^2(\xi_2^k-\theta^k)
         h(R_1,R_2)\xi_1'.\xi_2' W(\xi_1 {\circ}_\theta \xi_2)
.\eeq
 We will  able to simplify these relations  further.

Now we give the definition of the ``loop" product. First we  introduce
a concept of intersection of ``loops". Two ``loops" are said to intersect
if their $\theta^i$ coordinates coincide at some point. To make this more
precise, consider the given ``loops"
 $\xi_1:[0,L_1] \to \Sigma$  and $\xi_2 :[0,L_2] \to \Sigma$
,where we define $U=cst$ surface to be $\Sigma$.
They are said to intersect iff $\xi_1^k(l_1)=\xi_2^k(l_2)$
for some $l_1 \in [0,L_1]$ and $l_2 \in [0,L_2]$. One can
look at it somewhat more geometrically and say that two ``loops" intersect
if their projections to a transverse surface coincide somewhere (figure-2).
( This concept of intersection is the one implied by the delta functions in the
Poisson brackets.)
Next, we define
a product of two intersecting ``loops". Suppose that the point of intersection
is $Q_1$ on $\xi_1$ and $Q_2$ on $\xi_2$. Denote their corresponding
coordinates as $(R_1, \theta^i)$ and $(R_2, \theta^i)$ respectively.
Define $\xi_1 {\circ}_\theta \xi_2$ as the product with respect to
the pair $Q_1Q_2$ as follows: We start from $\xi_1(0)$ move along
$\xi_1$ till we reach the intersection point $Q_1$. At that point jump to
the ``loop" $\xi_2$ and start from the intersection point $Q_2$, move along
$\xi_2$ till $\xi_2(L_2)$. We jump down to $\xi_2(0)$ and move along
$\xi_2$ till we reach $Q_2$. We move back to $\xi_1$ where we left at $Q_1$
and continue along $\xi_1$ till we reach $\xi_1(L_1)$. Figure-3 clarifies
this product rule. Analytically we can express this in the following
formula;
\beq
       \xi_1:[0,L_1] \to \Sigma \quad \xi_2:[0,L_2] \to \Sigma
.\eeq
When they intersect, $\xi_1^i(l_1)=\xi_2^i(l_2)=\theta^i$,
\beq
       (\xi_1 {\circ}_\theta \xi_2)(k) =\cases{ \xi_1(k)  &for $k \in
                      [0,l_1]$\cr
                      \xi_2(k+(l_2-l_1)) &for $k \in
                      [l_1, l_1+L_2-l_2]$\cr
                     \xi_2(k-l_1-L_2+l_2) &for
                      $k \in [l_1+L_2-l_2, l_1+L_2]$\cr
                    \xi_1(k-L_2)  &for $k \in [l_1+L_2, L_1+L_2]$\cr}
.\eeq
This product is associative, i.e. it does not matter which way we
group the terms, if we keep the intersection pairs the same. This will
be usefull later on.

We will give the corresponding formulae in other dimensions
with a
Minkowski background; they are obtained exactly the same way, and the
definition of the ``loops" are also the same except this time the angular
variables are from $i=1$ to $d-2$ where $d=$dimension of the space-time.
The intersection and the product of Wilson ``loops" is also
defined analogously.
In the $U(N)$ case we  get,
\beqs{
 \{ W(\xi_1), W(\xi_2) \} =
              \int d\theta^1 d\theta^2.. d\theta^{d-2} ds_1 ds_2
    &\delta^{d-2}(\xi_1^k(s_1)-\theta^k) \delta^{d-2}
         (\xi_2^k(s_2)-\theta^k)\cr
        &f(r_1(s_1),r_2'(s_2)) \xi_1'.\xi_2' W(\xi_1 {\circ}_\theta \xi_2)
.}\eeqs
where $f(r,r')$ was defined before as an antisymmetric function.
One can also give the formula in the case where we use $r$ as a parameter.
In other gauge theories there will be corresponding quadratic factors. This
means that those algebras are not of Lie type.

A special property of four dimensional space--time is the two piece--wise
differentiable curves on a light--cone intersect generaically at a finite
number isolated points. In lower dimensions, the intersection will be along a
continuum, while in higher dimensions, generically there is no intersection at
all. For generic curves, the integrals in the Poisson algebra can be evaluated
to give  a sum,
\beq
      \{ W(\xi_1), W(\xi_2) \}= {\sum}_P h(R_1^P,R_2^P) {\xi_1'.\xi_2'
                               \over |\xi_1' \times \xi_2'| }{\Big |}_P
                                   W(\xi_1 \circ_P \xi_2)
.\eeq
The cross product is the usual antisymmetric
product of two vectors $\xi_1'$ and $\xi_2'$ without any angular factors.
The dot product is as
defined before, and $P$ refers to an intersection pair that we may also write
as $P{\bar P}$. The quantity ${\xi_1'.\xi_2'\over |\xi_1' \times \xi_2'| }{\Big
|}_P$ is ( upto a sign) the cotangent of the angle of intersection of the two
curves on $S^2$ at $P$. Also, $h(R_1,R_2)=1/2{\rm sign}(R_1-R_2)$.

The form of the algebra  shows explicitly
that the result is parameter independent and geometric in nature. In fact
it is possible to motivate this based on its very geometric character; this
will be the point of view we will pursue in the second part.

One should note the interesting situation that the $SU(N)$ case will lead to
a quadratic algebra of a fairly simple type. It can also be motivated
based on its purely geometric character. We state the result; it seems
that this is the only realistic situation where one has a quadratic algebra.

\beq
      \{ W(\xi_1), W(\xi_2) \}= {\sum}_P h(R_1^P,R_2^P) {\xi_1'.\xi_2'
                               \over |\xi_1' \times \xi_2'| }{\Big |}_P
                       (W(\xi_1 \circ_P \xi_2)-{1 \over N}W(\xi_1)W(\xi_2))
.\eeq

In this paper we will mainly concentrate on the $U(N)$ algebra and plan to
return to the quadratic case elsewhere.

 In three dimensions,
the theory is highly interacting, in the sense that the reduction to a
finite sum is not possible. This is due to the fact that
in three dimensions we need to use curves in two dimensions, and their
intersection is given by their projection to some transverse surface,
which is one dimensional in this case. But in one dimension any two ``loops"
intersect along a continuum. This renders the algebra to an integral
over the intersecting regions. One can work out the algebra in the generic
case. The situation in $d>4$
is generically simpler. We  see that the intersection
can be understood by projecting to a space of dimension$\geq 3$.
So it is very unlikely for ``loops" to intersect at all as the dimension
grows.  The Poisson algebra is generically trivial ( abelian).
 When the two ``loops" do
intersect the resulting expression is highly singular due to the
presence of the $\delta$-functions. In this case the result is not well-
defined.  This suggests that the ``right" dimension, so to speak,
 is four.

 In four dimensions there are also nongeneric
situation for which the algebra is not well-defined; such as two
``loops" intersecting along a continuous portion, or their tangent
vectors being paralel at the point of intersection. These we will
exclude by their nongeneric nature. To be more precise, we will
restrict ourselves to the class of ``loops" for which any pair of them
intersects at a finite number of points. This subset of ``loops" as we will
describe  in the second part , is going to be an
open dense subset of the space of ``loops". Our algebra is defined on this
subset of loops.
This is not such an unfamiliar
situation. In quantum mechanics for example most of the time we define
operators on an open dense subset, and try to find self-adjoint extensions
to a larger domain(if possible). We would like to have a similar philosophy
in this case. The details of these arguments are going to be given in the
second part, to avoid the repetition we will be somewhat heuristic in this
part.

 It is clear that the Wilson  `` loop" variables
 form  an overcomplete set of observables. Among all functions of $ \xi$,
those that represent parallel transport with respect to a   connection $A$ must
satisfy some constraints. Mandelstam obtained these constraints in the case of
based loops.  To obtain similar identities in our case, it will be convenient
to extend our ``loops" so that all of them pass through the points $i^-$ and
$i^0$ ( see figure-4).   They will then be based loops on compactified
Minkowsky space.
In fact we can attach a straight--line in the $R$ direction to the end--point
of any  ``loop" without changing the value of $W(A,\xi)$, since $A_R=0$.  Such
an extended ``loop" that goes from past infinity to spatial infinity will be
denoted typically by $\hat \xi$.

 Now, we can
state the  Mandelstam identities that the Wilson ``loop" variables must
satisfy\[mandelstam],\[giles].
 They simply capture the fact that $W$ is the trace of an $N\times N$
matrix.
For clarity we  first state  the
$U(1)$ and $U(2)$ gauge theories explicitly; for $U(1)$,
\beq
     W({\hat\xi}_1 \circ_P {\hat \xi}_2)=W({\hat \xi}_1) W({\hat \xi}_2)
.\eeq
where $P$ is any point of intersection of a pair of ``loops".

 For $U(2)$, the identity involves the simultaneous intersection of three
curves. Generically this will only happen at $i^-$ or $i^0$. At these points,
\beqs{
   &W({\hat \xi}_1 \circ_P {\hat \xi}_2 \circ_P {\hat \xi}_3)+
    W({\hat \xi}_1 \circ_P {\hat \xi}_3 \circ_P {\hat \xi}_2)-
    W({\hat \xi}_1 \circ_P {\hat \xi}_2)W({\hat \xi}_3)-
    W({\hat \xi}_1 \circ_P {\hat \xi}_3)W({\hat \xi}_2)-\cr
   &W({\hat \xi}_3 \circ_P {\hat \xi}_2)W({\hat \xi}_1)+
    W({\hat\xi}_1)W({\hat\xi}_2)W({\hat\xi}_3)=0
.}\eeqs
We used the associativity of the product and remove the paranthesis.

Now, we can state the $U(N)$ case,
\beqs{
     \sum_{\rm cycle \  decom. \ of \ S_{N+1}}
     & (-1)^\pi W({\hat \xi}_{i_1} \circ_P {\hat \xi}_{i_2}
                 \circ_P ...\circ_P {\hat \xi}_{i_{k_1}})
                W({\hat \xi}_{i_{k_1+1}} \circ_P {\hat \xi}_{i_{k_1+2}}
                 \circ_P ...\circ_P {\hat \xi}_{i_{k_1+k_2}})...\cr
         &W({\hat \xi}_{i_{k_1+k_2+...k_r+1}} \circ_P
                 {\hat \xi}_{i_{k_1+k_2+...k_r+2}}
                 \circ_P ...\circ_P {\hat \xi}_{i_{k_1+k_2+...k_{r+1}}})=0
.}\eeqs
where $k_1, k_2,....k_{r+1}$ denotes an arbitrary cycle. It satisfies
$k_1+k_2+...+k_{r+1}=N+1$ and the set $i_{k_1+...+k_s},...,i_{k_1+...+k_{s+1}}$
denotes a set of $k_{s+1}$ distinct numbers out of $1, 2,....N+1$.
A permutation is determined
through its  cycle decomposition,
and the above set of numbers specifies such
a cycle. Here $\pi$ is used to denote the sign of the permutation.
$S_{N+1}$ denotes the permutation group of $N+1$ numbers. $P$ is
also taken to be a common intersection point of $N=1$  ``loops"; and
generically $P$ can be only
$i^0$ or $i^-$.  A small deformation  of one of the curves will remove any
other simultaneous intersection. These are simplification of four dimensional
Yang--Mills theory.

 Dimension three
is exceptional, in the sense that there are an infinite
number of common intersection
points for an arbitrary set of ``loops". We can say that there are an
infinite number of Mandelstam identities in the case of three dimensions.
This means that there are an infinite number of constraints.
Therefore the theory could be highly interacting and complicated.

In the next section we will try to reverse the earlier point of view. We will
postulate a Lie bracket relation on the space of functions on loops. It will be
necessary to verify directly that these satisfy the Jacobi identity. Now it is
possible to define a dynamical system whose phase space is a co--adjoint orbit
of this  Lie algebra.  We will show that the Mandelstam identities define
surfaces in the co--adjoint orbits which are invariant under the action of the
Lie algebra.  Thus we suggest that the Phase spaces of $U(N)$ Yang--Mills
theories are the co--adjoint orbits of our Lie algebra. ( We are not able to
show this except for $U(1)$, since it is not clear if there are constraints
in addition to the Mandelstam constriants.) Eventually it might be possible to
derive Yang-Mills theory from a theory of loop variables, but we do not attempt
to do that here.

\sect{ The algebra of ``loops" in $\bar {\cal M}$ }

In this section we will study the ``loop" algebra {\it ab initio}: not
as derived from Yang-Mills theory.
Let us consider $\bar {\cal M}$ (it is described in section-3) and let $\Sigma$
be the $U={\rm constant}$ surface.

 Define ``loops"
on $\bar{\cal M}$ as follows. $\xi : [0,L] \to \Sigma $.  In particular we
choose $\Sigma$
to be ${\cal I}^-$ (since this the surface where we give the initial data).
We assume that $\xi^i(0)=\xi^i(L)$ where $i=1,2$ denotes the angular
coordinates. We do not require $\xi^R(0)$ to be equal to $\xi^R(L)$, hence
the term ``loop". These functions are assumed to be piecewise differentiable
and continuous only in the $\theta^k$ variables.

Next we introduce a concept of intersection for two ``loops" given by the
functions $\xi_1:[0,L_1] \to \Sigma$ and $\xi_2 :[0,L_2] \to \Sigma$.
Two ``loops" are
said to intersect if their angular coordinates coincide at some  pair of
points. More precisely, if $\xi_1^i(l_1)=\xi_2^i(l_2)$ for some $l_1 \in
[0,L_1] {\rm \ and \ } l_2 \in [0,L_2]$ and for $i=1,2$, then they
intersect at the corresponding points $Q_1(R_1, \theta^i)$ and $Q_2(R_2,
\theta^i)$ respectively. So an intersection is specified by a pair of
points (see figure-5). We also define a product of two intersecting ``loops".
The most natural one which will mix the geometry of two ``loops" is
given by the following rule; suppose that two ``loops" intersect at
the pair $Q_1Q_2$. Define $\xi_1 \circ_{Q_1Q_2} \xi_2 $ as the
product with respect to the  above pair. Start from $\xi_1(0)$ move along
it till we reach $Q_1$. At this point jump to the second ``loop" and
start from $Q_2$, move on $\xi_2$ till we reach $\xi_2(L_2)$. Then jump to
$\xi_2(0)$ and move along $\xi_2$ till we come to the intersection $Q_2$.
Now the motion along the second ``loop" is comlete, we jump back to the first
``loop" where we left it, namely the point $Q_1$. Continue to move on
$\xi_1$ till we reach $\xi_1(L_1)$. This gives the product loop. Since the
operation is piecewise differentiable the resulting object belongs to the
class of ``loops" that we specified. So algebraically the operation is closed.
The product can be given by the analytic expression;
\beq
       \xi_1:[0,L_1] \to \Sigma \quad \xi_2:[0,L_2] \to \Sigma
.\eeq
Assume that $\xi_1^i(l_1)=\xi_2^i(l_2)=\theta^i$,
\beq
       (\xi_1 {\circ}_\theta \xi_2)(k) =\cases{ \xi_1(k)  &for $k \in
                      [0,l_1]$\cr
                      \xi_2(k+(l_2-l_1)) &for $k \in
                      [l_1, l_1+L_2-l_2]$\cr
                     \xi_2(k-l_1-L_2+l_2) &for
                      $k \in [l_1+L_2-l_2, l_1+L_2]$\cr
                    \xi_1(k-L_2)  &for $k \in [l_1+L_2, L_1+L_2]$\cr}
.\eeq

Now, we introduce the radial extension of a ``loop". $\bar \xi$ is called
an extension of $\xi$ if $\bar \xi:[0,\bar L] \to \Sigma$
 and $\xi:[0,L] \to \Sigma$
such that $[0,L] \mapsto  [0,\bar L]$  by translation i.e. $x \in [0,L] \mapsto
x+a \in [0,\bar L]$ thus $[a, L+a] \subset [0,\bar L]$.
Furthermore, $\bar \xi|_{[a,
L+a]}=\xi$. This is the natural concept of extension. For our purposes, the
extensions
along the radial direction  are important. So, $\bar \xi$ is a radial extension
if $\bar\xi|_{[0,\bar L] \backslash [a,L+a]}$ remains a constant in $\theta^i$
directions(see figure-6). One should notice that our product rule
can be used to define the extension.

We denote the space of ``loops" considered as  directed geometric
objects by the set
$\cal L$ and define a function $W$ from this to complex numbers, {\it C}.
By that, we mean that the parametrization is essentially immaterial;
so in fact we
consider two ``loops" to be equivalent if they can be related by a
reparametrization of positive Jacobian. The direction is related to the
direction of time, since $\partial \over \partial R $ is a null vector,
the change of direction corresponds to the time reversal.
Thus, we require that $W$ does not depend
which particular representation has been chosen; it depends
 purely on the   intrinsic geometry and the direction.

Moreover, we postulate that $W(\xi)= W(\bar\xi)$ for any radial extension
$\bar \xi$. If we are looking at purely radial ``loops" we may introduce
an arbitrary normalization condition. This however is going to be fixed
by the constraints that we will introduce later on.
This will imply that the radial extension can be assumed to be completed
for any ``loop" as far as $W$ is concerned. This leads us to define the
maximal radial extension for a given ``loop". If we think of $\xi$
being  radially extended till the ends reach $i^0$ and $i^-$ respectively, the
resulting  ``loop" is called the maximal radial extension of $\xi$.
As we defined before there will be a representation of this in the
equivalence class of directed ``loops" $\cal L$.
We denote one such representative
as $\hat \xi$ and it satisfies $\hat \xi(0)=i^-$ and $\hat \xi(\hat L)=i^0$
or vice versa if $[0,\hat L]$ is the interval of parametrization.
It is enough to consider the maximally radially extended set of ``loops"
, therefore we denote this as $\hat {\cal L}$ and assume that $W$ is a
continuous function from $\hat {\cal L}$ to {\it C}. On the cartesian
product we introduce an algebra satisfied by $W$'s. It is postulated
as follows;
\beq
      \{ W(\hat\xi_1), W(\hat\xi_2) \}=
                       {\sum}_P h(R_1^P,R_2^{\bar P}) {\hat\xi_1'.\hat\xi_2'
                          \over |\hat\xi_1' \times \hat\xi_2'| }{\Big |}_P
                                   W(\hat\xi_1 \circ_P \hat\xi_2)
.\eeq
Here $h(R,R')=1/2{\rm sign}(R-R')$ and ${\hat \xi_1}'.{\hat \xi_2}'=\sum_k
{d{\hat \xi_1}^k \over ds_1}{d{\hat \xi_2}^k \over ds_2}\Phi^{-1} \Phi_k^2$
,${\hat\xi_1}' \times {\hat \xi_2}'={d{\hat \xi_1}^1 \over ds_1}
{d{\hat \xi_2}^2 \over ds_2}-
{d{\hat \xi_1}^2 \over ds_1}{d{\hat \xi_2}^1 \over ds_2}$. This has some
geometrical meaning. If we consider the tangent vectors their dot
product gives the cosine of the angle, and the denumerator is the
absolute value of the sine. There are some interesting points about
the conformal invariance of the algebra.[we need to expand this argument].
The notation $P,\bar P$ denotes the intersection pair; they have the
same angular coordinates but different $R$ values. This is the reason for using
$P$ for the angular part and distinguish the radial part.
For this algebra to make sense though we need to restrict ourselves to
a subset of $\hat {\cal L} \times \hat {\cal L}$. This is denoted by
$ \widehat {\cal {FL}}$ and defined to be the subset for which any pair of
``loops" have a finite number of intersections. This is in fact the generic
situation and the above subset is an open dense subset in
an appropriate topology. The extension of the
algebra  to the whole set can be defined by some extension process.

{}From now on, we consider ``loops" belonging to $\hat {\cal L}$ and the above
algebra postulated over the subset $\widehat {\cal{FL}}$.
We can state that the form of $h(R_1, R_2)$ guarantees the antisymmetry
of the algebra. Furthermore, by considering a generic situation it is
possible to prove the Jacobi identity on this subset(appendix-2). Namely,
\beq
    \{ W(\hat \xi_1),\{ W(\hat \xi_2) , W(\hat \xi_3) \} \} +
     {\rm cyclic \ perm. \ of\ (1,2,3)}=0
\eeq
The result is that the above algebra is a Poisson Algebra and it is
of the Lie type.

We want to introduce the Yang-Mills theory as a natural theory with the
phase space as the set of maximally radially extended curves. The
conjecture  here is that it should be possible to think of this as some
sort of an orbit structure. It is very similar to the coadjoint
orbits in Lie Groups. For symplectic action of a Lie group, we may
ask what  are the natural symplectic manifolds which admit this group
as a symmetry group. The answer is given by Krillov. The results are
the coadjoint orbits of the group with a natural symplectic form on them.
We do not have a group structure for the ``loops", but we have an algebraic
operation on them. On top of it we have a Poisson algebra of the Lie type.
We intrepret the action to be also generated by the Poisson algebra and
think of the resulting element as an element of the orbit. If we can give a
characterization of this orbit structure in another way this will be the
natural way to define a Gauge theory. We can indeed show that there
are some identities satisfied by $W$'s which are invariant under the
Poisson action. These will be some  constraint surfaces.

We introduce the constraint surfaces in the space of $W$'s. The most
natural set of constraints that we may introduce involves two ``loops" and
a simple quadratic relation. The next step would involve three ``loops"
and the most symmetric combination. We can go on like this and introduce
$N+1$ ``loops", which  satisfy a constraint  of order $N+1$ and involves
the most symmetric combination. These will be the Mandelstam identities.
For them to make sense, i.e. for ``loops" to be intermingle we expect
them to intersect. There are two such generic points, $i^-$ and $i^0$.
Other then that there will be some isolated set of configurations in the
$N+1$ fold product of $\hat {\cal L}$. This  will lead some reduction in the
degrees of freedom of the theory.

Let us state the Mandelstam identities; the simplest one will be,
\beq
     W({\hat\xi}_1 \circ_P {\hat \xi}_2)=W({\hat \xi}_1) W({\hat \xi}_2)
.\eeq
where $P$ is any point of intersection. Every pair  of ``loops" intersect
at $i^-$ and $i^0$.
\beqs{
   &W({\hat \xi}_1 \circ_P {\hat \xi}_2 \circ_P {\hat \xi}_3)+
    W({\hat \xi}_1 \circ_P {\hat \xi}_3 \circ_P {\hat \xi}_2)-
    W({\hat \xi}_1 \circ_P {\hat \xi}_2)W({\hat \xi}_3)-
    W({\hat \xi}_1 \circ_P {\hat \xi}_3)W({\hat \xi}_2)-\cr
   &W({\hat \xi}_3 \circ_P {\hat \xi}_2)W({\hat \xi}_1)+
    W({\hat\xi}_1)W({\hat\xi}_2)W({\hat\xi}_3)=0
.}\eeqs
where we take $P$ to be generically $i^0$ or $i^-$.

Therefore, the general such identity we can propose is given by,
\beqs{
      \sum_{\rm cycle \  decom. \ of \ S_{N+1}}
     & (-1)^\pi W({\hat \xi}_{i_1} \circ_P {\hat \xi}_{i_2}
                 \circ_P ...\circ_P {\hat \xi}_{i_{k_1}})
                W({\hat \xi}_{i_{k_1+1}} \circ_P {\hat \xi}_{i_{k_1+2}}
                 \circ_P ...\circ_P {\hat \xi}_{i_{k_1+k_2}})...\cr
         &W({\hat \xi}_{i_{k_1+k_2+...k_r+1}} \circ_P
                 {\hat \xi}_{i_{k_1+k_2+...k_r+2}}
                 \circ_P ...\circ_P {\hat \xi}_{i_{k_1+k_2+...k_{r+1}}})=0
.}\eeqs
where $k_1, k_2,....k_{r+1}$ denotes an arbitrary cycle. It satisfies
$k_1+k_2+...+k_{r+1}=N+1$ and the set $i_{k_1+...+k_s},...,i_{k_1+...+k_{s+1}}$
denotes a set of $k_{s+1}$ distinct numbers out of $1, 2,....N+1$.
A permutation is determined
through its  cycle decomposition,
and the above set of numbers specifies such
a cycle. Here $\pi$ is used to denote the sign of the permutation.
$S_{N+1}$ denotes the permutation group of $N+1$ numbers. $P$ is
also taken to be a common intersection point, and generically $P$ is
$i^0$ or $i^-$. There may be some other intersection points for
a set of $N+1$ ``loops", but obviously it is very unlikely even
for three of them to have such a common point.

We refer to the appendix-3 for the proof that these identities
are preserved under the Poisson action in  a generic case
Since the identities are preserved, it implies that the Poisson
action creates an orbit and each orbit is contained in one of
these surfaces. But the converse is being true is the conjecture;
that is,  the identities (with possibly some  restrictions,
such as $|W| \le 1$ ) are in fact enough to characterize the orbits.
This way we postulate that the above set of $W$'s on the space
$\widehat {\cal {FL}}$ with the above Poisson algebra when restricted
to a constraint surface of degree $N+1$ defines a $U(N)$ gauge theory.
Each such surface is in fact an orbit of some action generated by the
Poisson algebra itself(it is possible to give a more precise formulation).
We know that one representation of this
algebra is the four dimensional Yang-Mills theory in $\bar {\cal M}$
using the light-cone coordinates with the gauge choice $A_R=0$. The
general representation theory of such an algebra is of great interest.
Yet, we have no answer to such a question. The quantization of such
a theory is accomplished by finding irreducible representations in an
appropriate Hilbert space. This opens up a new set of questions
that we would like to look at. We can formally demostrate that
$U(1)$ gauge theory is determined by this set of identities if
we also add the pure phase condition, $W(\hat \xi)^*=W(\hat \xi)^{-1}$.
We can make the ansatz $W(\hat \xi)= e^{\phi(\hat \xi)}$
and put into the Poisson algebra;
\beq
    \{ W({\hat \xi}_1), W({\hat \xi}_2) \}= e^{\phi({\hat \xi}_1)}
     \{ \phi({\hat \xi}_1), \phi({\hat \xi}_2) \}  e^{\phi({\hat \xi}_2)}
.\eeq
But the other side is given by,
\beq
      \{ W(\hat\xi_1), W(\hat\xi_2) \}=
                       {\sum}_P h(R_1^P,R_2^P) {\hat\xi_1'.\hat\xi_2'
                          \over |\hat\xi_1' \times \hat\xi_2'| }{\Big |}_P
                                   W(\hat\xi_1 \circ_P \hat\xi_2)
.\eeq
Using the quadratic  constraint this can further be simplified to
\beq
      \{ W(\hat\xi_1), W(\hat\xi_2) \}=
                        F(\hat \xi_1, \hat \xi_2)
                             e^{\phi(\hat\xi_1)}e^{\phi (\hat\xi_2)}
.\eeq
where $F$ denotes just a geometric factor which depends on the two
``loops" and some intersection pairs. Cancelling out the common
factors, we see that the result is,
\beq
      \{ \phi({\hat \xi}_1), \phi({\hat \xi}_2) \}=
            F(\hat \xi_1, \hat \xi_2)
.\eeq
We can always bring the above constant matrix formally to diagonal form,
using an analogue of Darboux coordinates and in that the above function
is expressed as
\beq
      \phi(\hat \xi)=\int A_i d\hat \xi^i
.\eeq
for some constant
matrix (in the space of ``loops" it will be a matrix; ordinarily a
function in $\bar {\cal M}$ and it is a constant because it is not
a functional of $\xi$)
 $A_i$ and this is the desired representation.

\sect{ Appendix-1}

In this appendix we will give the derivation of the ``loop" algebra
assuming that the path ordered integral can be expressed via a piecewise
linear ``loop". In the limit as  the distances between these
and the original ``loop" go to zero in an appropriate
 topology  the result is going to be the integral.

Assume that $\eta$ is such a generic piecewise linear loop. We denote
the steps as $\Delta s$ and assume that $d(\eta,\xi) \mapsto 0$.
\beqs{
    W(\xi)=\lim_{\Delta \eta \mapsto 0} \Tr \prod (I+A_{i_k}\Delta \eta^{i_k})
.}\eeqs
where $\Delta \eta$ refers to the supremum over the interval sizes. We can
replace the intervals by the linear approximation and express this by
the parameter $s$ via $\Delta \eta=\eta'\Delta s$.
When we are calculating the Poisson brackets, we will implicitly assume
that the above limiting process is undertaken. For simplicity we keep the
same letter for the ``loop" integral.
Using the definition we can calculate the Poisson brackets of the
two ``loops".
\beqs{
      & \{ W(\xi_1), W(\xi_2) \}\cr
                      & =\lim_{ \Delta s_1, \Delta s_2 \mapsto 0}
            \{\Tr \prod_k (I+A_{i_k}\xi_1'^{i_k} \Delta s_{1,k}),
               \Tr \prod_l (I+A_{j_l}\xi_2'^{j_l} \Delta s_{2,l}) \}
.}\eeqs
We use the derivation property and express this as,

\beqs{
     \{ &W(\xi_1), W(\xi_2) \}=
     \lim_{ \Delta s_1, \Delta s_2 \mapsto 0} \sum_{m,n}
    \prod_{k < m} (I+A_{i_k}\xi_1'^{i_k}\Delta s_{1,k})^{\alpha_1}_{\alpha_2}
\prod_{l<n} (I+A_{j_l}\xi_2'^{j_l}\Delta s_{2,l})^{\alpha_6}_{\alpha_5} \cr
       &\{ A_{i_m,\alpha_3}^{\alpha_2}, A_{j_n,\alpha_4}^{\alpha_5} \}
         \xi_1'^{i_m} \xi_2'^{j_n}
           \Delta s_{1,m} \Delta s_{2,n}
       \prod_{k>m}  (I+A_{i_k}\xi_1'\Delta s_{1,k})^{\alpha_3}_{\alpha_1}
       \prod_{l>n}  (I+A_{j_l}\xi_2' \Delta s_{2,l})^{\alpha_4}_{\alpha_6}
.}\eeqs
where $\alpha$ has been used to denote the Lie Algebra contractions. We need to
employ the Poisson bracket of the $A$'s and also use the
completeness relation of the Lie Algebra generators.  This depends on the
way one will normalize the $U(1)$ generator and we give the
result for $U(N)$ with the large-$N$ convention and also $SU(N)$;

\beqs{
        &(T^a)^{\alpha_1}_{\beta_1}(T^a)^{\alpha_2}_{\beta_2}=
             \delta^{\alpha_1}_{\beta_2}\delta^{\alpha_2}_{\beta_1}\cr
        &(T^{a})^{\alpha_1}_{\beta_1}(T^a)^{\alpha_2}_{\beta_2}=
        \delta^{\alpha_1}_{\beta_2}\delta^{\alpha_2}_{\beta_1}
        -{1 \over N} \delta^{\alpha_1}_{\beta_1} \delta^{\alpha_2}_{\beta_2}
.}\eeqs
We normalized the generators of $SU(N)$ to $1$ and $I$ to $1/ \sqrt N$.
\beqs{
     \{ &W(\xi_1), W(\xi_2) \}=\cr
     &\lim_{ \Delta s_1, \Delta s_2 \mapsto 0} \sum_{m,n}
    \prod_{k < m} (I+A_{i_k}\xi_1'^{i_k}\Delta s_{1,k})^{\alpha_1}_{\alpha_2}
\prod_{l<n} (I+A_{j_l}\xi_2'^{j_l}\Delta s_{2,l})^{\alpha_6}_{\alpha_5}
       \delta_{\alpha_4}^{\alpha_2}\delta_{\alpha_3}^{\alpha_5}h(R_1^m,R_2^n)
              \Phi^{-1} \Phi^2_{i_m}\cr
        &\delta_{i_mj_n}
        \xi_1'^{i_m} \xi_2'^{j_n}\delta^2(\xi_1^{i_m}-\xi_2^{j_n})
           \Delta s_{1,m} \Delta s_{2,n}
       \prod_{k>m}  (I+A_{i_k}\xi_1'\Delta s_{1,k})^{\alpha_3}_{\alpha_1}
       \prod_{l>n}  (I+A_{j_l}\xi_2' \Delta s_{2,l})^{\alpha_4}_{\alpha_6}
.}\eeqs
If we used the $SU(N)$ expression there would be an extra term which did
not mix any internal indices. This is how one would get the quadratic term.
If we introduce the product $\xi_1'.\xi_2'=\sum_k \xi_1'^k \xi_2'^k \Phi^{-1}
\Phi^2_k$ and perform the Lie algebra contractions and rearrange the terms,
introduce an auxillary variable $\theta$ for the delta function,
we get,
\beqs{
     \{ &W(\xi_1), W(\xi_2) \}=\cr
     &\lim_{ \Delta s_1, \Delta s_2 \mapsto 0} \sum_{m,n}
    \prod_{k < m} (I+A_{i_k}\xi_1'^{i_k}\Delta s_{1,k})^{\alpha_1}_{\alpha_2}
\prod_{l<n} (I+A_{j_l}\xi_2'^{j_l}\Delta s_{2,l})^{\alpha_6}_{\alpha_3}
       h(R_1^m,R_2^n)\Delta s_{1,m} \Delta s_{2,n}   \cr
         &\int d\theta^1 d\theta^2 \xi_1'.
      \xi_2'|_{\theta}\delta^2(\xi_1^i(s_{1,m})
       -\theta^i) \delta^2(\xi_2^i(s_{2,n})-\theta^i)  \cr
   & \qquad \prod_{k>m}  (I+A_{i_k}\xi_1'\Delta s_{1,k})^{\alpha_3}_{\alpha_1}
       \prod_{l>n}  (I+A_{j_l}\xi_2' \Delta s_{2,l})^{\alpha_2}_{\alpha_6}
.}\eeqs
We rearrange and notice that,
\beqs{
     \lim_{ \Delta s_1, \Delta s_2 \mapsto 0}
 &\prod_{k < m} (I+A_{i_k}\xi_1'^{i_k}\Delta s_{1,k})^{\alpha_1}_{\alpha_2}\cr
   &\prod_{l>n}  (I+A_{j_l}\xi_2' \Delta s_{2,l})^{\alpha_2}_{\alpha_6}
\prod_{l<n} (I+A_{j_l}\xi_2'^{j_l}\Delta s_{2,l})^{\alpha_6}_{\alpha_3}
           \prod_{k>m}  (I+A_{i_k}\xi_1'\Delta s_{1,k})^{\alpha_3}_{\alpha_1}
.}\eeqs
is equal to $W(\xi_1 \circ_{\theta} \xi_2)$.
The remaining terms are nothing but integrals over the parameters;
\beqs{
     &\{ W(\xi_1), W(\xi_2) \} =\cr
             & {1\over N}\int d\theta^1 d\theta^2 ds_1 ds_2
        \delta^2(\xi_1^k(s_1)-\theta^k) \delta^2 (\xi_2^k(s_2)-\theta^k)
        h(R(s_1),R(s_2)) \xi_1'.\xi_2' W(\xi_1 {\circ}_\theta \xi_2)
.}\eeqs
The above derivation can easily be extended to higher dimensions.

\sect{ Appendix-2}

Here we give a proof of the Jacobi identity of the Poisson algebra, without
using the representation of $W$ in terms of Yang--Mills fields.
We will consider a generic situation as is shown in the figure-6. Every
``loop" has two intersection pairs with any other one. The generalization
to any other situation is clear except it is algebraically complicated.
We give the proof without refering to maximal radial extension, but this
does not change anything.

Let us look at the Poisson brackets of three ``loops";
\beqs{
       \{ W(\xi_1), \{ W(\xi_2), W(\xi_3) \} \} +
       \{ W(\xi_2), \{ W(\xi_3), W(\xi_1) \} \} +
       \{ W(\xi_3), \{ W(\xi_1), W(\xi_2) \} \}
.}\eeqs

We expand each of the Poisson brackets assuming the generic configuration
as in  figure-6. We use a condensed notation and drop the explicit
appearance of the angular factors in the algebra. Since they only depend
on the angular coordinates and they are symmetric functions we can
assume that they are attached to the functions $h(R,R')$. This will
simplify the writing of the expressions.
\beqs{
      & \{ W(\xi_1), h(R_2^{S_1},R_3^{\bar S_1})W(\xi_2 \circ_{S_1} \xi_3)+
                  h(R_2^{S_2},R_3^{\bar S_2})W(\xi_2 \circ_{S_2} \xi_3)\} =\cr
      & h(R_2^{S_1},R_3^{\bar S_1})[h(R_1^{Q_1},R_3^{\bar Q_1})
            W(\xi_1 \circ_{Q_1} (\xi_2 \circ_{S_1} \xi_3))+
            h(R_1^{\bar P_1},R_2^{P_1})
            W(\xi_1 \circ_{P_1} (\xi_2 \circ_{S_1} \xi_3))+\cr
      &     h(R_1^{\bar P_2},R_2^{P_2})
            W(\xi_1 \circ_{P_2} (\xi_2 \circ_{S_1} \xi_3))+
            h(R_1^{\bar Q_2},R_3^{Q_2})
            W(\xi_1 \circ_{Q_2} (\xi_2 \circ_{S_1} \xi_3))]+\cr
      & h(R_2^{S_2},R_3^{\bar S_2})[h(R_1^{Q_1},R_3^{\bar Q_1})
            W(\xi_1 \circ_{Q_1} (\xi_2 \circ_{S_2} \xi_3))+
            h(R_1^{\bar P_1},R_2^{P_1})
            W(\xi_1 \circ_{P_1} (\xi_2 \circ_{S_2} \xi_3))+\cr
      &     h(R_1^{\bar P_2},R_2^{P_2})
            W(\xi_1 \circ_{P_2} (\xi_2 \circ_{S_2} \xi_3))+
            h(R_1^{\bar Q_2},R_3^{Q_2})
            W(\xi_1 \circ_{Q_2} (\xi_2 \circ_{S_1} \xi_3))]
.}\eeqs
Similarly for the others,
\beqs{
      & \{ W(\xi_2), h(R_3^{\bar Q_1},R_1^{Q_1})W(\xi_3 \circ_{Q_1} \xi_1)+
                  h(R_3^{Q_2},R_1^{\bar Q_2})W(\xi_3 \circ_{Q_2} \xi_1)\} =\cr
      & h(R_2^{\bar Q_1},R_1^{Q_1})[h(R_2^{S_1},R_3^{\bar S_1})
            W(\xi_2 \circ_{S_1} (\xi_3 \circ_{Q_1} \xi_1))+
            h(R_2^{P_1},R_1^{\bar P_1})
            W(\xi_2 \circ_{P_1} (\xi_3 \circ_{Q_1} \xi_1))+\cr
      &     h(R_2^{P_2},R_1^{\bar P_2})
            W(\xi_2 \circ_{P_2} (\xi_3 \circ_{Q_1} \xi_1))+
            h(R_2^{S_2},R_3^{\bar S_2})
            W(\xi_2 \circ_{S_2} (\xi_3 \circ_{Q_1} \xi_1))]+\cr
      & h(R_3^{Q_2},R_1^{\bar Q_2})[h(R_2^{S_1},R_3^{\bar S_1})
            W(\xi_2 \circ_{S_1} (\xi_3 \circ_{Q_2} \xi_1))+
            h(R_2^{P_1},R_1^{\bar P_1})
            W(\xi_2 \circ_{P_1} (\xi_3 \circ_{Q_2} \xi_1))+\cr
      &     h(R_2^{P_2},R_1^{\bar P_2})
            W(\xi_2 \circ_{P_2} (\xi_3 \circ_{Q_2} \xi_1))+
            h(R_2^{S_2},R_3^{\bar S_2})
            W(\xi_1 \circ_{S_2} (\xi_3 \circ_{Q_2} \xi_1))]
.}\eeqs
\beqs{
      & \{ W(\xi_3), h(R_1^{\bar P_1},R_2^{P_1})W(\xi_1 \circ_{P_1} \xi_2)+
                  h(R_1^{\bar P_2},R_2^{P_2})W(\xi_1 \circ_{P_2} \xi_2) \}=\cr
      & h(R_1^{\bar P_1},R_2^{P_1})[h(R_3^{\bar S_1},R_1^{S_1})
            W(\xi_3 \circ_{S_1} (\xi_1 \circ_{P_1} \xi_2))+
            h(R_3^{\bar Q_1},R_1^{Q_1})
            W(\xi_3 \circ_{Q_1} (\xi_1 \circ_{P_1} \xi_2))+\cr
      &     h(R_3^{Q_2},R_1^{\bar Q_2})
            W(\xi_3 \circ_{Q_2} (\xi_1 \circ_{P_1} \xi_2))+
            h(R_3^{\bar S_2},R_2^{S_2})
            W(\xi_3 \circ_{S_2} (\xi_1 \circ_{P_1} \xi_2))]+\cr
      & h(R_1^{\bar P_2},R_2^{P_2})[h(R_3^{\bar S_1},R_1^{S_1})
            W(\xi_3 \circ_{S_1} (\xi_1 \circ_{P_2} \xi_2))+
            h(R_3^{\bar Q_1},R_1^{Q_1})
            W(\xi_3 \circ_{Q_1} (\xi_1 \circ_{P_2} \xi_2))+\cr
      &     h(R_3^{Q_2},R_1^{\bar Q_2})
            W(\xi_3 \circ_{Q_2} (\xi_1 \circ_{P_2} \xi_2))+
            h(R_3^{\bar S_2},R_2^{S_2})
            W(\xi_3 \circ_{S_2} (\xi_1 \circ_{P_2} \xi_2))]
.}\eeqs

Let us add up these terms and combine the related ones;
For $W(\xi_1 \circ_{Q_1} (\xi_2 \circ_{S_1} \xi_3))$
\beqs{
       h(R_2^{S_1},R_3^{\bar S_1})h(R_1^{Q_1},R_3^{\bar Q_1})+
       h(R_3^{\bar Q_1},R_1^{Q_1})h(R_2^{S_1},R_3^{\bar S_1})=0
.}\eeqs
For $W(\xi_1 \circ_{P_1} (\xi_2 \circ_{S_1} \xi_3))$
\beqs{
       h(R_2^{S_1},R_3^{\bar S_1})h(R_1^{P_1},R_2^{\bar P_1})+
       h(R_1^{\bar P_1},R_2^{P_1})h(R_3^{\bar S_1},R_2^{S_1})=0
.}\eeqs
For $W(\xi_1 \circ_{P_2} (\xi_2 \circ_{S_1} \xi_3))$
\beqs{
       h(R_2^{S_1},R_3^{\bar S_1})h(R_1^{\bar P_2},R_2^{P_2})+
       h(R_1^{\bar P_2},R_2^{P_2})h(R_3^{\bar S_1},R_2^{S_1})=0
.}\eeqs
For $W(\xi_1 \circ_{Q_2} (\xi_2 \circ_{S_1} \xi_3))$
\beqs{
       h(R_3^{\bar Q_2},R_1^{Q_2})h(R_2^{S_1},R_3^{\bar S_1})+
       h(R_2^{S_1},R_3^{\bar S_1}) h(R_1^{Q_2},R_3^{\bar Q_2})=0
.}\eeqs
For $W(\xi_1 \circ_{Q_1} (\xi_2 \circ_{S_2} \xi_3))$
\beqs{
       h(R_2^{S_2}, R_3^{\bar S_2})h(R_1^{Q_1}, R_3^{\bar Q_1})+
       h(R_3^{\bar Q_1}, R_1^{Q_1}) h(R_2^{S_2}, R_3^{\bar S_2})=0
.}\eeqs
For $W(\xi_1 \circ_{P_1} (\xi_2 \circ_{S_2} \xi_3))$
\beqs{
       h(R_2^{S_2},R_3^{\bar S_2})h(R_1^{\bar P_1},R_2^{P_1})+
       h(R_1^{\bar P_1},R_2^{P_1})h(R_3^{\bar S_2}, R_2^{S_2})=0
.}\eeqs
For $W(\xi_1 \circ_{P_2} (\xi_2 \circ_{S_2} \xi_3))$
\beqs{
       h(R_2^{S_2},R_3^{\bar S_2})h(R_1^{\bar P_2}, R_2^{P_2})+
       h(R_1^{\bar P_2}, R_2^{P_2})h(R_3^{\bar S_2},R_2^{S_2})=0
.}\eeqs
For $W(\xi_1 \circ_{Q_2} (\xi_2 \circ_{S_2} \xi_3))$
\beqs{
       h(R_2^{S_2},R_3^{\bar S_2})h(R_1^{\bar Q_2},R_3^{Q_2})+
       h(R_3^{Q_2},R_1^{\bar Q_2})h(R_2^{S_2},R_3^{\bar S_2})=0
.}\eeqs
For $W(\xi_2 \circ_{P_1}(\xi_3 \circ_{Q_1} \xi_1))$
\beqs{
       h(R_3^{\bar Q_1},R_1^{Q_1})h(R_2^{P_1},R_1^{\bar P_1})+
       h(R_1^{\bar P_1},R_2^{P_1}) h(R_3^{\bar Q_1},R_1^{Q_1})=0
.}\eeqs
For $W(\xi_2 \circ_{P_2}(\xi_3 \circ_{Q_1} \xi_1))$
\beqs{
       h(R_3^{\bar Q_1},R_1^{Q_1})h(R_2^{P_2},R_1^{\bar P_2})+
       h(R_2^{P_2},R_1^{\bar P_2})h(R_1^{Q_1},R_3^{\bar Q_1})=0
.}\eeqs
For $W(\xi_2 \circ_{P_1} (\xi_3 \circ_{Q_2} \xi_1))$
\beqs{
       h(R_3^{Q_2},R_1^{\bar Q_2})h(R_2^{P_1},R_1^{\bar P_1})+
       h(R_1^{\bar P_1}, R_2^{P_1})h(R_3^{Q_2},R_1^{\bar Q_2})=0
.}\eeqs
For $W(\xi_2 \circ_{P_2}(\xi_3 \circ_{Q_2} \xi_1))$
\beqs{
      h(R_3^{Q_2},R_1^{\bar Q_2})h(R_2^{P_2},R_1^{\bar P_2})+
      h(R_1^{\bar P_2},R_2^{P_2})h(R_3^{Q_2},R_1^{\bar Q_2})=0
.}\eeqs

Using antisymmetry of the $h(R,R')$ function we get all the above
terms zero and essentially even if we do not have the figure, the
antisymmetry gives zero. We also used the fact that paranthesis
can be removed from products in the calculation of $W$'s keeping
the order and product pair the same. One should also note that
$W(\xi_1 \circ_{Q} \xi_2)=W(\xi_2 \circ_{Q} \xi_1)$ since interchange
can be taken as a reparametrization with a discontinuity; and we allow
these type of finite discontinuities of the  parameter.

\sect{Appendix-3}

In this case we will consider $U(2)$ Mandelstam constraint-
which  is a cubic  relation. We will prove that it is invariant under
the action of the Poisson algebra.
 We will consider a generic case and take a nongeneric
intersection just to show that the argument is essentially independent of
the way the intersection is assumed. This is shown in figure-7.
The generalization to other cases is
straightforward.

Let us call the triple intersection points $Q_1Q_2Q_3$ and denote this by
$Q$ collectively.

\beqs{
   &W( \xi_1 \circ_Q \xi_2 \circ_Q \xi_3)+
    W(\xi_1 \circ_Q  \xi_3 \circ_Q  \xi_2)-
    W(\xi_1 \circ_Q \xi_2)W(\xi_3)-
    W(\xi_1 \circ_Q \xi_3)W(\xi_2)-\cr
   &W(\xi_3 \circ_Q \xi_2)W(\xi_1)+
    W(\xi_1)W(\xi_2)W(\xi_3)=0
.}\eeqs
Next, we calculate the Poisson bracket of this with $\xi_4$ and
show that the right hand side and the left hand side both give zero.
Here is the expansion of each term, we are again using condensed notation
of keeping angular factor as part of the antisymmetric function $h(R,R')$;

\beqs{
      &\{ W(\xi_4),W(\xi_1)W(\xi_2)W(\xi_3) \}=\cr
      &[h(R_4^{P_1},R_1^{\bar P_1})W(\xi_4 \circ_{P_1} \xi_1)+
   h(R_4^{\bar P_2},R_1^{P_2})W(\xi_4 \circ_{P_2} \xi_1)]W(\xi_2)W(\xi_3)+\cr
     &[h(R_4^{\bar S_1},R_2^{S_1})W(\xi_4 \circ_{S_1} \xi_2)+
   h(R_4^{\bar S_2},R_2^{S_2})W(\xi_4 \circ_{S_2} \xi_2)]W(\xi_1)W(\xi_3)+\cr
     &[h(R_4^{T_2},R_3^{\bar T_2})W(\xi_4 \circ_{T_2} \xi_3)+
   h(R_4^{\bar T_1},R_3^{T_1})W(\xi_4 \circ_{T_1} \xi_3)]W(\xi_1)W(\xi_2)
.}\eeqs

\beqs{
     &- \{ W(\xi_4), W(\xi_1 \circ_Q \xi_2)W(\xi_3)+
                   W(\xi_2 \circ_Q \xi_3)W(\xi_1)+
                   W(\xi_3 \circ_Q \xi_1)W(\xi_2) \}=\cr
      &-[h(R_4^{P_1},R_1^{\bar P_1})W(\xi_4 \circ_{P_1} \xi_1 \circ_Q \xi_2)
     +h(R_4^{\bar S_1},R_2^{S_1})W(\xi_4 \circ_{S_1} \xi_1 \circ_Q \xi_2)+\cr
      &h(R_4^{\bar S_2},R_2^{S_2})W(\xi_4 \circ_{S_2} \xi_1 \circ_Q \xi_2)+
       h(R_4^{\bar P_2},R_1^{P_2})W(\xi_4 \circ_{P_2} \xi_1 \circ_Q \xi_2)]
        W(\xi_3)\cr
      &-[h(R_4^{T_2},R_3^{\bar T_2})W(\xi_4 \circ_{T_2} \xi_3)+
         h(R_4^{\bar T_1},R_3^{T_1})W(\xi_4 \circ_{T_1} \xi_3)]
        W(\xi_1 \circ_Q \xi_2)\cr
      &-[h(R_4^{\bar S_1},R_2^{S_1})W(\xi_4 \circ_{S_1} \xi_2 \circ_Q \xi_3)
     +h(R_4^{T_2},R_3^{\bar T_2})W(\xi_4 \circ_{T_2} \xi_2 \circ_Q \xi_3)+\cr
      &h(R_4^{\bar T_1},R_3^{T_1})W(\xi_4 \circ_{T_1} \xi_2 \circ_Q \xi_3)+
      h(R_4^{\bar S_2},R_2^{S_2}) W(\xi_4 \circ_{S_2} \xi_2 \circ_Q \xi_3)]
         W(\xi_1)\cr
      &-[h(R_4^{P_1},R_1^{\bar P_1})W(\xi_4 \circ_{P_1} \xi_1)+
         h(R_4^{\bar P_2}, R_1^{P_2})W(\xi_4 \circ_{P_2} \xi_1)]
         W(\xi_2 \circ_Q \xi_3)\cr
      &-[h(R_4^{P_1},R_1^{\bar P_1})W(\xi_4 \circ_{P_1} \xi_3 \circ_Q \xi_1)+
      h(R_4^{\bar P_2},R_1^{P_2})W(\xi_4 \circ_{P_2} \xi_3 \circ_Q \xi_1)+\cr
      &h(R_4^{T_2},R_3^{\bar T_2})W(\xi_4 \circ_{T_2} \xi_3 \circ_Q \xi_1)+
       h(R_4^{\bar T_1},R_3^{T_1})W(\xi_4 \circ_{T_1} \xi_3 \circ_Q \xi_1)]
         W(\xi_2)\cr
      &-[h(R_4^{\bar S_1},R_2^{S_1})W(\xi_4 \circ_{S_1} \xi_2)+
         h(R_4^{\bar S_2},R_2^{S_2})W(\xi_4 \circ_{S_2} \xi_2)]
        W(\xi_3 \circ_Q \xi_1)
.}\eeqs
\beqs{
       &\{ W(\xi_4),W(\xi_1 \circ_Q \xi_2 \circ_Q \xi_3)+
                   W(\xi_1 \circ_Q \xi_3 \circ_Q \xi_2) \}=\cr
       &[h(R_4^{P_1},R_1^{\bar P_1})W(\xi_4 \circ_{P_1} \xi_1 \circ_Q
         \xi_2 \circ_Q \xi_3)+
         h(R_4^{\bar S_1},R_2^{S_1})W(\xi_4 \circ_{S_1} \xi_1 \circ_Q
         \xi_2 \circ_Q \xi_3)+\cr
       &h(R_4^{\bar P_2},R_1^{P_2})W(\xi_4 \circ_{P_2} \xi_1 \circ_Q
         \xi_2 \circ_Q \xi_3)+
        h(R_4^{T_2},R_3^{\bar T_2})W(\xi_4 \circ_{T_2} \xi_1 \circ_Q
         \xi_2 \circ_Q \xi_3)+\cr
        & h(R_4^{\bar T_1},R_3^{T_1})W(\xi_4 \circ_{T_1} \xi_1 \circ_Q
         \xi_2 \circ_Q \xi_3)+
          h(R_4^{\bar S_2},R_2^{S_2})W(\xi_4 \circ_{S_2} \xi_1 \circ_Q
         \xi_2 \circ_Q \xi_3)]+\cr
       &[h(R_4^{P_1},R_1^{\bar P_1})W(\xi_4 \circ_{P_1} \xi_1 \circ_Q
         \xi_3 \circ_Q \xi_2)+
         h(R_4^{\bar S_1},R_2^{S_1})W(\xi_4 \circ_{S_1} \xi_1 \circ_Q
         \xi_3 \circ_Q \xi_2)+\cr
       &h(R_4^{\bar P_2},R_1^{P_2})W(\xi_4 \circ_{P_2} \xi_1 \circ_Q
         \xi_3 \circ_Q \xi_2)+
        h(R_4^{T_2},R_3^{\bar T_2})W(\xi_4 \circ_{T_2} \xi_1 \circ_Q
         \xi_3 \circ_Q \xi_2)+\cr
        & h(R_4^{\bar T_1},R_3^{T_1})W(\xi_4 \circ_{T_1} \xi_1 \circ_Q
         \xi_3 \circ_Q \xi_2)+
          h(R_4^{\bar S_2},R_2^{S_2})W(\xi_4 \circ_{S_2} \xi_1 \circ_Q
         \xi_3 \circ_Q \xi_2)]
.}\eeqs
We can combine term by term. The main idea is that the addition of $\xi_4$
to each ``loop" is going to create another ``loop" with the same common
point of intersection and it has to satisfy the Mandelstam constraint.
This gives zero for each such term. We will combine the terms with
the coefficient $h(R_4^{P_1},R_1^{\bar P_1})$;
\beqs{
     h(R_4^{P_1},R_1^{\bar P_1})&[W(\xi_4 \circ_{P_1} \xi_1)W(\xi_2)W(\xi_1)
       -W(\xi_4 \circ_{P_1} \xi_1 \circ_Q \xi_2)W(\xi_3)\cr
      &-W(\xi_4 \circ_{P_1} \xi_1 \circ_Q \xi_3)W(\xi_2)
       -W(\xi_4 \circ_{P_1} \xi_1)W(\xi_2 \circ_Q \xi_3)+\cr
      &W(\xi_4 \circ_{P_1} \xi_1 \circ_Q \xi_2 \circ_Q \xi_3)+
       W(\xi_4 \circ_{P_1} \xi_1 \circ_Q \xi_3 \circ_Q \xi_2)]
.}\eeqs
and the above term is zero because of the Mandelstam constraint for
$(\xi_4 \circ_{P_1} \xi_1),\xi_2,\xi_3$ with respect to the intersection
$Q$. We also used the fact that the angular factors attached to them
depends only which curve we are on, but not to the specific product.

{\bf Acknowledgements}
We would like to thank A. Ashtekar,S. Guruswamy, R. Henderson and K. Gupta
for usefull
discussions. Alan Fry kindly helped to prepare the figures.
This work was supported in part by the US Department of Energy, Grant
No. DE-FG02-91ER40685.

{\bf References}\hfill\break

\noindent\mandelstam. S. Mandelstam, Phys. Rev. {\bf 175} 1580 (1968);
                                     Phys. Rev. {\bf D19} 2391 (1979).

\noindent\rajeev. S.G.Rajeev, Phys. Lett. {\bf B212} 203 (1988).

\noindent\turgutetal. K.G. Gupta, R.J. Henderson, S.G. Rajeev, and
O.T. Turgut, Jour. Math. Phys. {\bf 35 } 3845 (1994)

\noindent\chandar. L. Chandar and E. Ercolessi, Nucl. Phys. {\bf B426}
 94 (1994).

\noindent\trieste. S.G. Rajeev,   in Proceedings of the 1991 Summer School in
High Energy Physics and Cosmology, ed. E. Gava et. al. World Scientific (1992)
p. 955.

\noindent\twodbaryon. P. Bedaque, I. Horvath and S. G. Rajeev, Mod. Phys. Lett.
{\bf A7} 3347 (1992).

\noindent\twodhadron. S.G. Rajeev, hepth \#9401115
 to appear in Int. Jour. Mod. Phys.

\noindent\pesando. I. Pesando, hep-th 9408018; M. Caviddi, P. di Vecchia and
I. Pesando, Mod. Phys. Lett. {\bf A8} 2127(1993).

\noindent\wadia. A. Dhar, G. Mandal and S. R. Wadia,
    Phys. Lett. {\bf B329} 15 (1994); hep-th 9407026.

\noindent\sphqcd. K.S. Gupta, S. Guruswamy, and S.G. Rajeev,
Phys. Rev. {\bf D48} 3354 (1993).

\noindent\migdal. M. Makeenko and A. A. Migdal, Nucl. Phys. {\bf B188}
269 (1981).

\noindent\giles. R. Giles, Phys. Rev {\bf D24} 2160 (1981).

\noindent\kikkawa. K. Kikkawa, Ann. Phys. {\bf 135} 222 (1981).

\noindent\ashtekaretal. A. Ashtekar,{\it Lectures on Non-perturbative
Canonical Gravity}, Singapore, World Scientific (1991);
A. Ashtekar and C. Isham,
Class. Quan. Grav. {\bf 9} 1433 (1993).

\noindent\annphys. S.G. Rajeev, Ann. Phys. {\bf 173} 249 (1987).

\noindent\lightcone. R. J. Perry, Ann. Phys. {\bf 232} 116 (1994).

\noindent\drinfeld. V. G. Drin'feld, {\it Quantum Groups}, ICM Proceedings,
Berkeley, 898 (1986).

\noindent\kakukikkawa. M. Kaku and K. Kikkawa, Phys. Rev. {\bf D10} 1110
(1974); Phys. Rev. {\bf D10} 1823 (1974).

\noindent\penrose. R.Penrose and W. Rindler, {\it Spinors and Space-time},
Cambridge  Univ. Press, 1986; R. Wald, {\it General Relativity},
Univ. of Chicago Press, 1984.

\noindent\thooft. G. t'Hooft, Nucl. Phys. {\bf B72} 461 (1974).

\bye